# Size-induced acoustic hardening and optic softening of phonons in CdS, InP, CeO$_2$, SnO$_2$, and Si nanostructures

Chang Q. Sun,[*] L. K. Pan, C. M. Li, and S. Li[#]
School of Electrical and Electronic Engineering Nanyang Technological University Singapore 639798
#School of Materials Science and Engineering, The University of New South Wales, Sydney NSW 2052, Australia

It has been puzzling that the Raman optical modes shift to lower frequency (or termed as optical mode softening) associated with creation of Raman acoustic modes that shift to higher energy (or called as acoustic hardening) upon nanosolid formation and size reduction. Understandings of the mechanism behind the size-induced acoustic hardening and optic softening have been quite controversial. On the basis of the recent bond order-length-strength (BOLS) correlation [Phys. Rev. B 69 045105 (2004)], we show that the optical softening arises from atomic cohesive energy weakening of surface atoms and the acoustic mode hardening is predominated by intergrain interaction. Agreement between predictions and observations has been reached for Si, CdS, InP, TiO$_2$, CeO$_2$, and SnO$_2$ nanostructures with elucidation of vibration frequency of the corresponding isolated dimers. Findings further evidence the impact of bond order loss to low-dimensional systems and the essentiality of the BOLS correlation in describing the behavior of nanostructures.

E-mail ecqsun@ntu.edu.sg Fax 65 6792 0415 URL www.ntu.edu.sg/home/ecqsun/

I  Introduction

Atomic vibration is of high interest because the behavior of phonons influence directly on the electrical and optical properties in solid materials and devices such as electron-phonon coupling in photoabsorption and photoemission, and phonon scattering in device transport dynamics.[1] It has been long surprising that with structural miniaturization down to nanometer scale the transverse and the longitudinal optical (TO/LO) Raman modes shift towards lower frequency (or called as optical mode softening)[2] accompanied with generation of low-frequency Raman (LFR) acoustic modes at wave numbers of a few or a few tens cm$^{-1}$. The LFR peak shifts up (or called as acoustic mode hardening) towards higher frequency upon the solid size being reduced.[3,4] Generally, the size dependent Raman shifts follow a scaling relation:[2,4]

$$\omega(K_j) = \omega(\infty) + A_f / K_j^{\kappa}$$

where $A_f$ and $\kappa$ are adjustable parameters for data fitting. $K_j$, the dimensionless form of size, is the number of atoms with diameter $d$ lined along the radius ($R_j$) of a spherical dot. For optical red shift, $A_f < 0$. For Si example, $\omega(\infty) = 520$ cm$^{-1}$ corresponds to wavelength of $2\times10^4$ nm and the index $\kappa$ varies from 1.08 to = 1.44 or even 2.0, varying from source to source.[5] For the LFR blue shift, $A_f > 0$, $\kappa = 1$, and $\omega(\infty) = 0$. Therefore, the LFR disappears for large particles.

   The underlying mechanism behind the Raman shift is under debate with numerous theories. Theoretical studies of phonon frequency shift are often based on continuum dielectric mechanism.[6,7] Sophisticated calculations have been carried out using models of correlation length,[8] bulk phonon dispersion,[9] lattice-dynamic matrix,[10] associated with microscopic valence force field,[4] phonon confinement,[11] and bond polarization.[2]

   The mechanism of quadrupolar vibration taking the individual nanoparticle as a whole was assumed to be responsible for the LFR acoustic modes. The phonon energies are size dependent and vary with materials of the host matrix. The LFR scattering from silver nanoclusters embedded in porous alumina[12] and SiO$_2$[13] was suggested to arise from the quadrupolar vibration modes that are enhanced by the excitation of the surface plasmas of the encapsulated Ag particles. The selection of modes by LFR scattering is suggested to arise from the stronger plasmon-phonon coupling for these modes. For an Ag particle smaller than four nanometers, the size dependence of the peak frequency can be explained using Lamb's theory[14] that gives vibrational frequencies of a homogeneous elastic body with a spherical form. On the other hand, lattice strain was suggested to be another possible mechanism for the LFR blue shift as size-dependent compressive strain has been observed from CdS$_x$Se$_{1-x}$ nanocrystals embedded in a borosilicate (B$_2$O$_3$-SiO$_2$) glass matrix.[15] The lattice strain enhances the surface stress when the crystal size is reduced. Therefore, the observed blue shift of acoustic phonon energies was suggested to be consequence of the compressive stress that overcomes the red shift caused by phonon confinement. Liang et al[16] presented a model for the Raman blue shift by relating the frequency shift to the bond length and bond strength that are functions of entropy latent heat of fusion and the critical temperature for solid-liquid transition.

The high-frequency Raman shift has ever been suggested to be activated by surface disorder,[17] surface stress,[18,19] and phonon quantum confinement,[20,21] as well as surface chemical passivation.[22] The phonon confinement model attributes the red shift of the Raman line to the relaxation of the wave-vector selection rule ($\Delta q = 0$) for the excitation of the Raman active phonons due to their localization. The relaxation of the selection rule arises from the finite crystalline size and the diameter distribution of nanosolid in the films. When the size is decreased the rule of momentum conservation will be relaxed and the Raman active modes will not be limited at the center of the Brillouin zone.[18] A Gaussian-type phonon confinement model[21] indicates that strong phonon damping presents whereas calculations[23] using the correlation functions of the local dielectric constant ignores the role of phonon damping in the nanosolid. The large surface-to-volume ratio of a nanodot strongly affects the optical properties because of introducing surface polarization and surface states.[24] Using a phenomenological Gaussian envelope function of phonon amplitudes, Tanaka et al.[25] showed that the size dependence of optic red shift originated from the relaxation of the Δq = 0 selection rule based on the phonon confinement argument with negative phonon dispersion. The phonon energies for all the glasses are reduced and the values of the phonon energies of CdSe nanodots are found to be quite different for different host glasses. A sophisticated analytical model of Hwang et al.[5] indicates that the effect of lattice strain must be considered in explaining the optical red shift for CdSe nanodots embedded in different glass matrices. For a free surface, it has been derived that the red shift follows the relation:

$$\frac{\Delta\omega(K_j)}{\omega(\infty)} = BK_j^{-2}$$

(1)

The value of $B$ in eq (1) is a competition between the phonon negative dispersion and the size-dependent surface tension. Thus, a positive value of $B$ indicates that the phonon negative dispersion exceeds the size-dependent surface tension and consequently causes the red shift of phonon frequency, and vice versa. In case of balance of the two effects, i.e. $B = 0$, the size dependence disappears. There are still some difficulties to use this equation, as remarked by Hwang et al.[5]

It is noted that currently available models for the optical red shift are based on assumptions that the materials are homogeneous and isotropic which is valid only in the long-wavelength limit. When the size of the nanosolid is in the range of a few nanometers the continuum dielectric models exhibit limitations. Therefore, the discussed models could hardly reproduce with satisfactory the Raman frequency shifts at the lower end of the size limit. The objective of this work is to show that derivatives of the recent BOLS correlation mechanism[26,27,28] could reproduce the size induced Raman shifts leading to deeper and consistent insight into the mechanism behind with information about the vibration frequency of the corresponding dimers, which is beyond the scope of other sophisticated models.

II Principle
2.1 Vibration modes

Raman scattering is known to arise from the radiating dipole moment induced in a system by the electric field of incident electromagnetic radiation. The laws of momentum and energy conservation govern the interaction between a phonon and the incident photon. When we consider a solid containing numerous Bravais unit cells, and each cell contains $n$ atoms, there will be $3n$ modes of vibrations. Among the $3n$ modes, there will be three acoustic modes, LA, $TA_1$, and $TA_2$, and 3(n-1) optical modes, LO and TOs. The acoustic modes represent the in-phase motion of the mass center of the unit cell or the entire solid as a whole. The long-range Coulomb interaction is responsible for the intercluster interaction. Therefore, the acoustic LFR should arise from the vibration of the entire nanosolid interacting with the host matrix or with other neighboring clusters. Therefore, it is expected that the LTR mode approaches zero if the particle size is infinitely large. The optical modes arise from the relative motion of the individual atoms in a complex unit cell. For elemental solids with a simple crystal structure such as the fcc of Ag, only acoustic modes present. Silicon or diamond is an interlock of two fcc unit cells that contain each cell two atoms in nonequivalent positions, there will be three acoustic modes and three optical modes.

2.2 Optical phonon frequency

The total energy $E$ causing lattice vibration consists of the component of short-range interactions $E_S$ and the component of long-range Coulomb interaction $E_C$[4]

$$E = E_S + E_C. \tag{2}$$

The long-range part corresponds to the LFR mode and represents the weak interaction between nanosolids. The short-range energy $E_S$ arising from nearest bonding atoms, which is composed of two parts. One is the lattice thermal vibration $E_V(T)$ and the other is interatomic binding energy at zero K, $E_b(r)$. The $E_S$ for a dimer can be expressed in a Taylor's series,[26]

$$\begin{aligned} E_S(r,T) &= \sum_n \left( \frac{d^n u(r)}{n! dr^n} \right)_{r=d} (r-d)^n \\ &= u(d) + 0 + \left.\frac{d^2 u(r)}{2! dr^2}\right|_d (r-d)^2 + \left.\frac{d^3 u(r)}{3! dr^3}\right|_d (r-d)^3 \ldots \\ &= E_b(d) + \frac{k}{2}(r-d)^2 + \frac{k'}{6}(r-d)^3 + \ldots \\ &= E_b(d) + E_V(T) \end{aligned} \tag{3}$$

The term with index $n = 0$ corresponds to the minimal binding energy at T = 0 K, $E_b(d) < 0$. The term $n = 1$ is the force [ $\partial u(r)/\partial r|_d = 0$] at equilibrium and the terms $n \geq 2$ correspond to the thermal vibration energy, $E_V(T)$. By definition, the thermal vibration energy of a single bond is

$$\begin{aligned} E_V(T) &= \mu \omega^2 (r-d)^2 / 2 = k_v (r-d)^2 / 2 \\ &= \sum_{n \geq 2} \left( \frac{d^n u(r)}{n! dr^n} (r-d)^{n-2} \right)_d (r-d)^2 \end{aligned} \tag{4}$$

where $r-d = x$ is the magnitude of lattice vibration. $\mu$ is the reduced mass of the dimer of concern. The $k_v = \mu \omega^2 \propto E_b/d^2$ is the force constant for lattice harmonic vibration with an angular frequency of $\omega$. High-order terms correspond to nonlinear contribution that can be negligible in the first order approximation.

For a single bond, the $k_v$ is strengthened because of the bond order loss induced bond contraction and bond strength gain.[26-29] For a single atom, we have to count contribution from all the neighboring bonds. For a lower-coordinated atom the resultant $k_v$ could be lower because of the bond order loss. Considering the vibration amplitude $x \ll d$, it is convenient and reasonable to take the mean contribution from each coordinate to the force constant and to the magnitude of dislocation as the first order approximation:

$$k_1 = k_2 = \cdots = k_z = \mu_i \omega^2$$

and $\quad x_1 = x_2 = \cdots = x_z = (r-d)/z$.

Therefore the total energy of a certain atom with $z$ coordinates is the sum over all the coordinates,

$$E_S(d,T) = \sum_z \left[ E_b + \frac{\mu \omega^2}{2}\left(\frac{r-d}{z}\right)^2 + \ldots \right]$$

$$= zE_b + \frac{zd^2u(r)}{2!\,dr^2}\bigg|_d (r-d)^2 + \ldots$$

(5)

This relation leads to the expression for phonon frequency as a function of bond energy and atomic CN, and bond length,

$$\omega = z \times \left[\frac{d^2u(r)}{\mu dr^2}\bigg|_d\right]^{\frac{1}{2}} \propto \frac{zE_b^{1/2}}{d}$$

(6)

According to the BOLS correlation,[26-29] the bond order loss of a surface atom causes the remaining bonds of the lower-coordinated atoms to contract spontaneously ($d_i = c_i d$) associated with bond strength gain ($E_i = c_i^{-m} E_b$). The index $m$ recognizes the nature of the bond involved. Such a BOLS correlation and its consequence modify not only the atomic cohesive energy (atomic CN multiplies the single bond energy) but also the Hamiltonain due to the densification of binding energy in the relaxed region. A physically detectable quantity that depends on the atomic cohesive energy or the Hamiltonian for a nanosolid can be expressed as $Q(K_j)$ in a shell structure:

$$\begin{cases} Q(K_j) = Nq_0 + N_S(q_S - q_0) \\ \dfrac{Q(K_j) - Q(\infty)}{Q(\infty)} = \sum_{i \le 3} \gamma_i \dfrac{\Delta q}{q} \end{cases}$$

(7)

where $Q(\infty) = Nq_0$ is for a bulk solid. $q_0$ and $q_S$ correspond to the $Q$ value per atomic volume inside the bulk and in the surface region, respectively. $N_S = \Sigma N_i$ is the number of atoms in the surface atomic shells. Combining eqs. (6) and (7) gives the size-dependent optic red shift (where $Q(\infty) = \omega(\infty) - \omega(1)$):

$$\frac{\omega(R) - \omega(\infty)}{\omega(\infty) - \omega(1)} = \sum_{i \le 3} \gamma_i \left[\frac{\omega_i}{\omega_b} - 1\right] = \sum_{i \le 3} \gamma_i \left[\frac{z_i}{z_b} c_i^{-\left(\frac{m}{2}+1\right)} - 1\right] = \Delta_p < 0$$

where

$$\begin{cases} \gamma_i = \dfrac{N_i}{N} = \dfrac{V_i}{V} \approx \begin{cases} 1, & K_j \le 3 \\ \dfrac{\tau c_i}{K_j}, & else \end{cases} \\ c_i = 2/\{1 + \exp[(12 - z_i)/(8 z_i)]\} \\ z_1 = 4(1 - 0.75/K_j) \quad (spherical) \\ z_2 = 6;\ z_3 = 12 \end{cases}$$

(8)

$\omega(1)$ is the vibrational frequency of an isolated dimer which is the reference point for the optical red shift upon nanosolid and bulk formation. $\gamma_i$ is the portion of atoms in the $i$th atomic layer over the total number of atoms of the entire solid of different shapes($\tau = 1$–3 correspond to a thin plate, a rod, and a spherical dot, respectively). The index $i$ is counted up to three from the outmost atomic layer to the center of the solid as no atomic CN imperfection is justified at i > 3.

III  Results and discussion
3.1 Optical modes and dimer vibration

In experiment, one can only measure $\omega(\infty)$ and $\omega(K_j)$ in eq (8). However, with the known $m$ value derived from measurement of other quantities such as the melting point or core level energy,[26-29] one can determine $\omega(1)$ or the bulk shift $\omega(\infty) - \omega(1)$ by matching the measured data represented below to the predicted line in eq **(8)** without needing any other assumptions,

$$\Delta\omega(K_j) = \begin{cases} \dfrac{-A'}{K_j^{\kappa}}, & (Measurement) \\ = \Delta_R[\omega(\infty) - \omega(1)], & (Theory) \end{cases}$$

(9)

Hence, the frequency shift from the dimer bond vibration to the bulk value, $\omega(\infty) - \omega(1) \equiv -A'/(\Delta_R K_j^{\kappa})$, can be obtained. The matching of prediction with measurement indicates that $k \equiv 1$, because $\Delta_R \propto K_j^{-1}$.

Figure 1 shows that the BOLS predictions match exceedingly well with the theoretically calculated or the experimentally measured optical red shift of a number of samples. Derived information about the corresponding dimer vibration is given in Table **1**.

3.2    Acoustic modes and intercluster interaction
Figure 2 shows the least-square-mean-root fitting of the size dependent LFR frequency for different nanosolids. The LFR frequency depends linearly on the inverse $K_j$

$$\omega(K_j) - \omega(\infty) = \dfrac{-A'}{K_j}$$

**(10)**

The zero intercept at the vertical axis, $\omega(\infty) = 0$, indicates that when the $K_j$ approaches infinity the LFR peaks disappear, which implies that the LFR modes and their blue shifts originate from vibration of the individual nanoparticle as a whole. It seems not essential to involve the quadruple motion or the bond strain at the interface. However, the current derivative gives information about the strength of interparticle interaction, as summarized in Table 2.

3.3 Surface atom vibration

According to Einstein's relation, it can be derived that $\mu(c\omega x)^2/2z = k_B T$. At a given temperature, the vibrational amplitude and frequency of a given atom is correlated as: $x \propto z^{1/2}\omega^{-1}$, which is CN dependent. The frequency and magnitude of vibration for an surface atom at the surface (z = 4) or a metallic monatomic chain (MC with z = 2) can be derived as

$$\frac{\omega_1}{\omega_b} = z_{ib}c_1^{-(m/2+1)} = \begin{cases} 0.88^{-3.44}/3 = 0.517 & (Si, m = 4.88) \\ 0.88^{-3/2}/3 = 0.404 & (Metal, m = 1) \\ 0.70^{-3/2}/6 = 0.2846 & (MC, m = 1) \end{cases}$$

and

$$\frac{x_1}{x_b} = (z_1/z_b)^{1/2}\,\omega_b/\omega_1 = (z_b/z_1)^{1/2}c_1^{(m/2+1)}$$

$$= \begin{cases} \sqrt{3} \times 0.88^{3.44} = 1.09 & (Si) \\ \sqrt{3} \times 0.88^{3/2} = 1.43 & (Metal) \\ \sqrt{6} \times 0.70^{3/2} = 1.43 & (MC) \end{cases}$$

(11)

The vibrational amplitude of an atom at the surface or a MC is indeed greater than that of a bulk atom while the frequency is lower. The magnitude and frequency are sensitive to the *m* value and varies insignificantly with the curvature of a spherical dot when $K_j > 3$. This result verifies for the first time the assumption[30,31] that the vibration amplitude of a surface atom is always greater than the bulk value and it keeps constant at all particle sizes.

IV Summary

In summary, a combination of the BOLS correlation and the scaling relation has enabled us to correlate the size-created and the size-hardened LFR acoustic phonons to the intergrain interaction and the optic phonon softening to the CN-imperfection reduced cohesive energy of atoms near the surface edge. The optic softening and acoustic hardening is realized in a $K_j^{-1}$ fashion. Decoding the measured size-dependence of Raman optical shift has derived vibrational information of Si, InP, CdS, CdSe, $TiO_2$, $CeO_2$, and $SnO_2$ dimers and their bulk shifts, which is beyond the scope of direct measurement. As the approach proceeds in a way from bond-by-bond, atom-by-atom, shell-by-shell, no other constraints developed for the continuum medium are applied. One striking significance is that we are able to verify the correlation between the magnitude and the frequency of vibration of the lower-coordinated atoms. Consistency between the BOLS predictions and observations also verify the validity of other possible models that incorporate the size-induced Raman shift from different perspectives.

Table and Figure captions

Table 1 Vibration frequencies of isolated dimers of various nanosolids and their red shift upon bulk formation derived from simulating the size dependent red shift of Raman optical modes as shown in Figure 1.

| Material | d (nm) | A′ | $\omega(\infty)$ (cm$^{-1}$) | $\omega(1)$ (cm$^{-1}$) | $\omega(\infty)-\omega(1)$ (cm$^{-1}$) |
|---|---|---|---|---|---|
| $CdS_{0.65}Se_{0.35}$ | 0.286 | 23.9 | 203.4 | 158.8 | 44.6 |
|  | 0.286 | 24.3 | 303 | 257.7 | 45.3 |
| CdSe | 0.294 | 7.76 | 210 | 195.2 | 14.8 |
| $CeO_2$ | 0.22 | 20.89 | 464.5 | 415.1 | 49.4 |
| $SnO_2$ | 0.202 | 14.11 | 638 | 602.4 | 35.6 |
| InP | 0.294 | 7.06 | 347 | 333.5 | 13.5 |
| Si | 0.2632 | 5.32 | 520.0 | 502.3 | 17.7 |

Table 2 Linearization of the LFR acoustic modes of various nanosolids gives information about the strength of interparticle interaction for the specific solids.

| Sample | A′ |
|---|---|
| Ag-a & Ag-b | 23.6 ± 0.7 |
| Ag-c | 18.2 ± 0.6 |
| $TiO_2$-a $TiO_2$-b | 105.5 ± 0.1 |
| $SnO_2$-a | 93.5 ± 5.4 |
| CdSe-1-a | 146.1 ± 6.27 |
| CdSe-1-b | 83.8 ± 2.8 |
| CdSe-1-c | 46.7 ± 1.4 |
| CdSSe-a | 129.4 ± 1.2 |
| CdSSe-b | 58.4 ± 0.8 |
| Si-LA | 97.77 |
| Si-$TA_1$ | 45.57 |
| Si-$TA_2$ | 33.78 |

Figure 1 ([link](link)) Comparison of the BOLS predictions (lines for different shapes) with theoretical and experimental observations (scattered data) on the size-dependent optic phonon softening of nano-solid. (a) data labeled Si-1 was calculated using correlation length model,[8] Si-3 (dot) and Si-4 (rod) were calculated using the bulk dispersion relation of phonons,[9] Si-5 was calculated from the lattice-dynamic matrix,[4] Si-7 was calculated using phonon confinement model,[11] and Si-8 (rod) and Si-9 (dot) were calculated using bond polarizability model.[2] Data for Si-2,[32] Si-6,[33] and Si-10, and Si-11[18] are measured data. (b) $CdS_{0.65}Se_{0.35}$-1, $CdS_{0.65}Se_{0.35}$ (in glass)-$LO_2$, $CdS_{0.65}Se_{0.35}$-2, $CdS_{0.65}Se_{0.35}$ (in

glass)-LO$_1$,[34] CdSe-1, CdSe(in B$_2$O$_3$SiO$_2$)-LO, CdSe-2, CdSe(in SiO$_2$)-LO, and CdSe-3 CdSe(in GeO$_2$)-LO, CdSe-4, CdSe(in GeO$_2$)-LO,[25] (c) CeO$_2$-1,[35] SnO$_2$-1,[36] SnO$_2$-2,[17] and InP[37] are all measurement.

Figure 2 ([link]) Generation and blue shift of the LFR acoustic modes where the solid dotted and dashed lines are the corresponding results of the least squares fitting. (a) the Si-a, Si-b, and Si-c were calculated from the lattice-dynamic matrix by using a microscopic valence force field model,[4] the Si-d and Si-e are the experimental results.[5] (b) Ag-a (Ag in SiO$_2$)[38] Ag-b (Ag in SiO$_2$)[13] Ag-c (Ag in Alumina).[12] (c) TiO$_2$-a[39] TiO$_2$-b[39] SnO$_2$-a.[17] (d) CdSe-a (l = 0 n = 2) CdSe-b (l = 2 n = 1) and CdSe-c (l = 0 n = 1).[40] (e) CdS$_{0.65}$Se$_{0.35}$-a [CdS$_{0.65}$Se$_{0.35}$ (in glass)-LF2] and CdS$_{0.65}$Se$_{0.35}$-b [CdS$_{0.65}$Se$_{0.35}$ (in glass)-LF1][34] are all measured data.

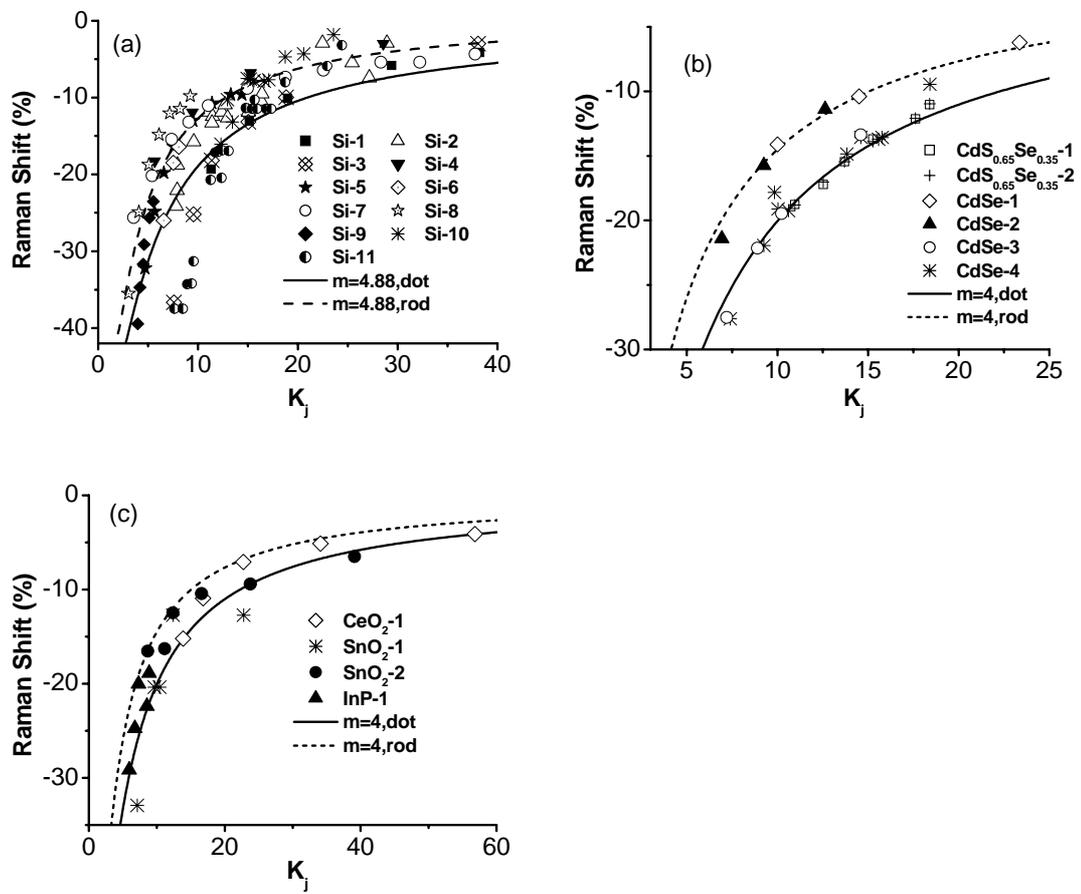

Fg-1

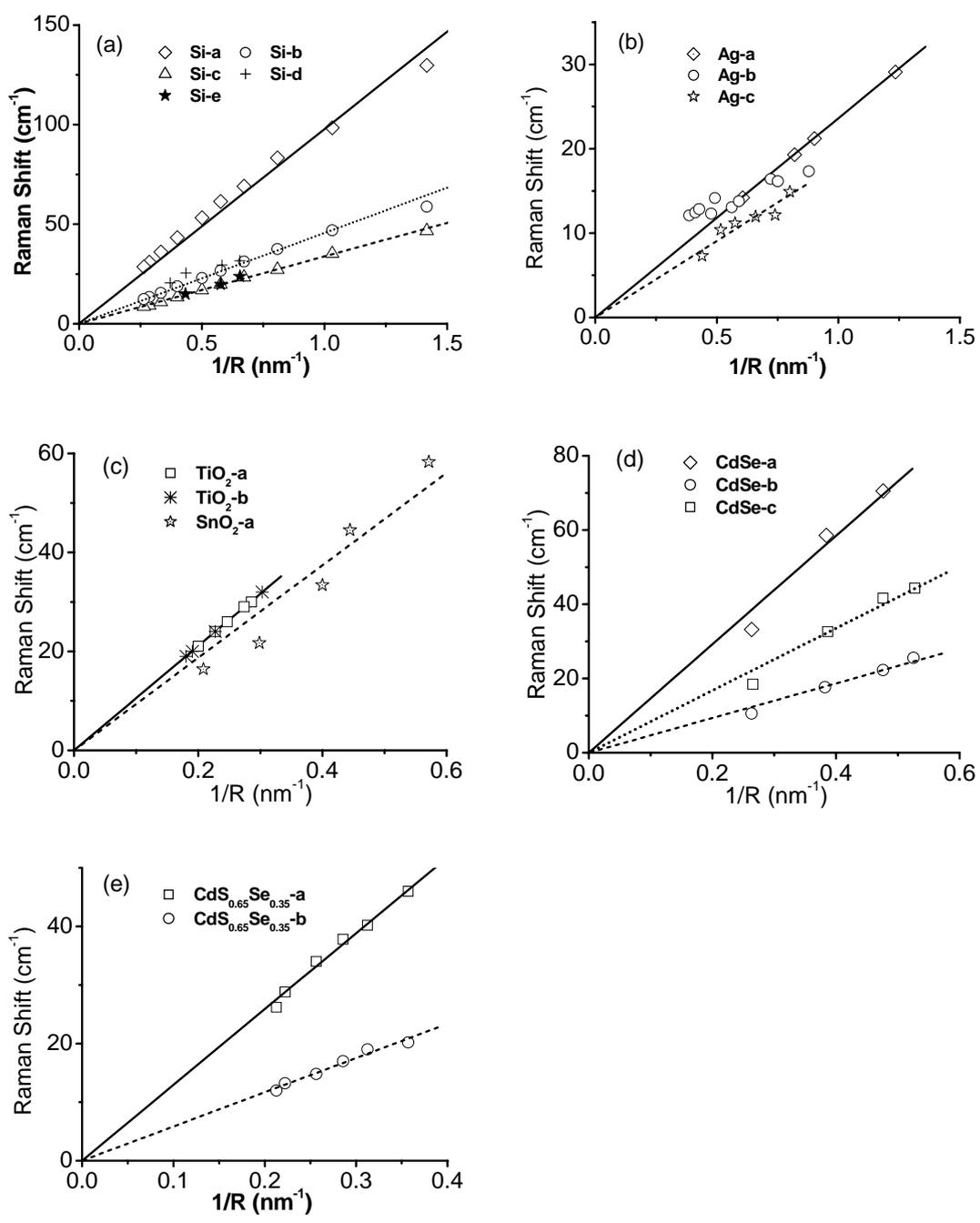

Fg-2